\documentclass[aps,prl,floatfix,twocolumn,amsmath,amssymb]{revtex4-1}
\usepackage{amsthm,color,latexsym,array,epstopdf}
\usepackage[hyperfootnotes=true,breaklinks]{hyperref}
\usepackage{graphicx}
\usepackage{grffile}

\usepackage{feynmp-auto}
\usepackage{bbm,slashed,tikz}
\definecolor{darkred}{rgb}{0.5,0.0,0.0}
\definecolor{darkblue}{rgb}{0.0,0.0,0.9}
\definecolor{darkerblue}{rgb}{0.0,0.0,0.5}
\definecolor{darkgreen}{rgb}{0.0,0.5,0.0}
\definecolor{black}{rgb}{0.0,0.0,0.0}
\definecolor{brown}{rgb}{0.6,0.4,0.2}

\renewcommand{\Re}{\operatorname{Re}}

\def\be{\begin{equation}}
    \def\ee{\end{equation}}

\def\cO{\mathcal{O}}
\def\cI{\mathcal{I}}

\def\Li{\text{Li}}

\begin{document}

    \begin{fmffile}{feyngraph}
        \unitlength = 1mm
        
        \title{The  Compton Scattering Total Cross Section at Next-to-Leading Order}
        \author{Roman N. Lee}
        \email{r.n.lee@inp.nsk.su}
        \affiliation{Budker Institute of Nuclear Physics, 630090, Novosibirsk, Russia}
        \author{Matthew D.~Schwartz}
        \email{schwartz@g.harvard.edu}
        \affiliation{Department of Physics, Harvard University, Cambridge, MA 02138}
        \author{Xiaoyuan Zhang}
        \email{xiaoyuanzhang@g.harvard.edu}
        \affiliation{Department of Physics, Harvard University, Cambridge, MA 02138}

        \begin{abstract}
            An analytic formula is given for the total scattering cross section of an electron and a photon at order $\alpha^3$. This includes both the double-Compton scattering real-emission contribution as well as the virtual Compton scattering part. When combined with the recent analytic result for the  pair-production cross section, the complete $\alpha^3$ cross section is now known. Both the next-to-leading order calculation as well as the pair-production cross section are computed using modern multiloop calculation techniques, where cut diagrams are decomposed into a set of master integrals that are then computed using differential equations.
        \end{abstract}
        
        \maketitle
        
        The scattering of photons off of electrons is perhaps {\it the} most important physical processes in nature: essentially all observable phenomena involve photon-electron interactions. The shift in the wavelength of the scattered photon, $\Delta \lambda = \frac{h}{m c} (1-\cos\theta)$, as first observed by Compton in 1923 convincingly demonstrated that light comprises particles with energy and momentum~\cite{PhysRev.21.483}. Compton's paper introduced the Compton wavelength of the electron, $\lambda_e = \frac{h}{m c}$, which governs the effective cross sectional area of the electron as seen by the photon $\sigma \sim \pi \lambda_e^2$. The calculation of the
        Compton scattering cross section by Dirac~\cite{dirac1926relativity} and Gordon~\cite{gordon1926comptoneffekt}, and later with full spin and relativistic corrections by Klein and Nishina~\cite{klein1929scatter}, provided a convincing case of the correctness of the Dirac equation. Photon scattering off of electrons is critical to a wide variety of scientific enterprise, from X-ray crystallography to astrophysics. 
        
        Total cross sections at high energy are of interest for both experimental reasons and theoretical ones. On the experimental side, they are relevant for not just applications like cosmic rays, but also for estimating luminosity and measuring coupling constants. On theoretical side, total cross sections necessarily involve the singular forward-scattering region, where outgoing particles are collinear to incoming ones. In this region, off-shell Glauber/Coulomb modes are essential. Studying these modes has led to insights such as Regge physics and the BFKL equation~\cite{Lipatov:1976zz,Kuraev:1976ge,Balitsky:1978ic}. For Compton scattering, the singularities are in a region where the outgoing electron is collinear to the incoming photon and dominated by $t$-channel fermion exchange. Despite their importance, very few analytic results are known for total cross sections beyond the leading order. 
        
        Bethe and Heitler~\cite{Bethe:1934za}, as well as Racah~\cite{Racah1934a}, considered a related process of pair production in a background electromagnetic field by the high-energy photon, $\gamma Z\to Z e^+e^-$. This process has a total cross section which scales asymptotically like $m^{-2}\ln s$, rather than $s^{-1}$, an indication of the relevance of the Glauber region. Upon formal substitution $Z=-1$, the leading high-energy asymptotics of this cross section also determines the high-energy limit of $e^- \gamma \to e^- e^+ e^-$ process (cf. Ref. \cite{baier1966electromagnetic}).   
        The complete analytical result for the total cross section of the latter was first computed only recently~\cite{Lee:2020obg} (using the same technology developed for this paper), confirming the leading high-energy asymptotics of Bethe and Heitler. The analytic $e^+e^- \to \gamma \gamma$ cross section at NLO has also been completed within the last year by the same methods~\cite{Lee:2020zpo}. The cross section of $\gamma \gamma\to e^+e^-$ can, in principle, be extracted as the Abelian limit of the $gg \to t\bar{t}$ total cross section~\cite{Czakon:2008ii}. Compton scattering is thus the last pure QED total cross section not known analytically at order $\alpha^3$ in QED. Even the leading high-energy asymptotics of the cross section is not known.
        
        The importance of Compton scattering led numerous investigators to explore corrections beyond the leading order.
        The next-to-leading order (NLO) Compton scattering process includes both loop contributions to $e^- \gamma \to e^-\gamma$ and real double-Compton scattering $e^- \gamma \to e^- \gamma \gamma$. These two contributions are separately infrared divergent, but the divergence cancels when they are added, as guaranteed by the Bloch-Nordsieck theorem~\cite{Bloch:1937pw}. The virtual graphs were computed by Feynman and Brown in 1951~\cite{Brown:1952eu}, and the cancellation of the infrared divergence in the differential cross section was shown using a photon mass cutoff. The double Compton process was studied by Mandl and Skyrme in 1952~\cite{mandl1952theory}. Recently, the total cross section for double Compton scattering has been calculated in Ref. \cite{Lee:2020obg}. The asymptotic behavior of Compton scattering at high energy at the amplitude level has been examined by numerous authors (e.g. ~\cite{gell1964elementary,mccoy1976theory,sen1983asymptotic}).
        Polarized differential Compton scattering at NLO was studied  by Swartz~\cite{Swartz:1997im} and by Denner and Dittmaier~\cite{Denner:1998nk}. Although numerical results for the total NLO cross section can be obtained by integrating these differential cross sections over the scattering angle, no analytic formula has yet been produced. The result of this paper is that final missing analytic form.
        
\begin{figure*}[t!]
            \centering
            \includegraphics[]{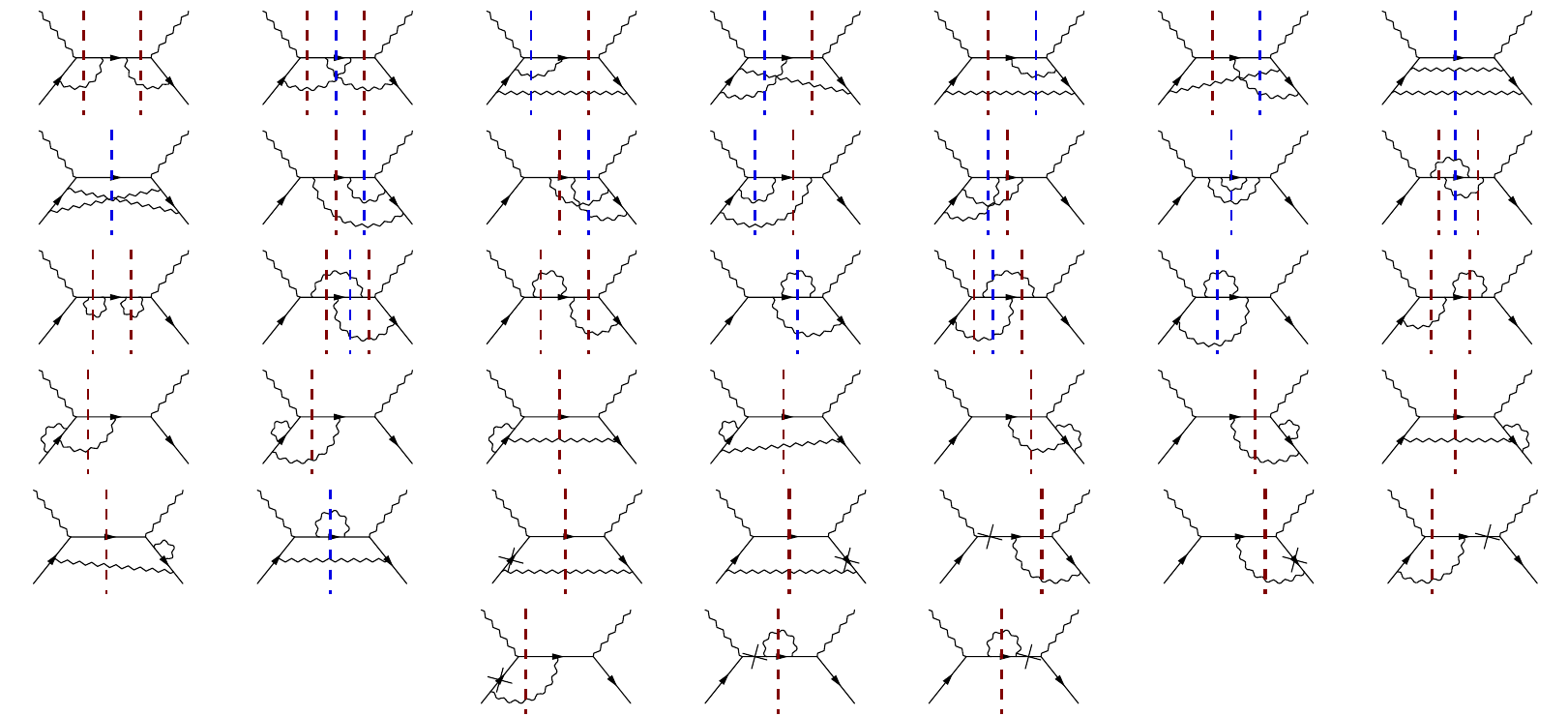}
            \caption{Cut Feynman diagrams contributing to the NLO cross section. Cuts drawn in blue indicate double-Compton contributions and red cuts indicate loop contributions. The last eight diagrams involve an insertion of the mass counterterm and require separate integrals.}
        \label{fig:diagrams}
        \end{figure*}

        To compute the total cross section, one approach is to use the optical theorem
        %: $\sigma_\text{tot} = 2 \text{Im}(\cM_F)$ where $\cM_F$ 
        to extract it from the imaginary part of the $e^- \gamma \to e^- \gamma$ forward scattering amplitude.
However, rather than compute the full 2-loop forward amplitude and then take its imaginary part, it is simpler to compute the cut diagrams directly. These diagrams are shown in Fig.~\ref{fig:diagrams}. The cuts that put $e^-\gamma$ on-shell are the virtual corrections, while those putting $e^-\gamma\gamma$ on-shell correspond to real emission. There are also contributions to the total $e^- \gamma$ cross section at order $\alpha^3$ from final states with 3 charged particles. These were computed in~\cite{Lee:2020obg} so we do not consider them here.
        
        To compute the cuts, we apply integration by parts and differential equations to 2 or 3 particle cuts separately. 
        For example, one of 2-particle cut master integrals of interest looks like
        \begin{multline}\label{eq:2cut}
\hspace{-3mm}
\cI_1 = 
            \begin{gathered}
            \includegraphics[]{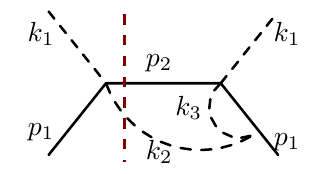}
            \end{gathered}\\
            =
            \frac{(2 \pi )^2}{2 i \pi ^d}\int d^d p_2 d^d k_2 d^d k_3
            \delta(p_2^2-m^2)\theta(p_2^0)
             \delta(k_2^2)\theta(k_2^0)
            \\
            \times \frac{1}{k_3^2[ (p_1-k_2-k_3)^2-m^2]}
            \delta^d(p_1+k_1-p_2-k_2)
        \end{multline}
        Applying loop computation technology to cut graphs significantly simplifies the problem: the extra $\delta$-functions reduce the number of master integrals and, therefore, the size of the differential system. Even more important, the cuts relevant for the NLO correction to the Compton scattering cross section prevent the appearance of the non-polylogarithmic master integrals --- the massive sunrise graphs. 
        
        The main tool we use for the IBP (integration-by-parts) reduction is \texttt{LiteRed} \cite{Lee2013a,Lee2021} which allows for the account of the individual cuts. For the reduction of the differential system to $\epsilon$-form we use \texttt{Libra} package \cite{lee2020libra}.
        It is helpful to rewrite the integrals in terms of threshold variables like
        \begin{equation}
            x = \frac{s-m^2}{m^2},
        \quad
            y=\sqrt{\frac{x}{x+4}}\,. \label{xy}
        \end{equation}
        A variable like $y$ can be used to rationalize the weights of the appearing iterated integrals. Both $x$ and $y$ vanish at threshold $s \to m^2$. We find that using the threshold limit to set the boundary conditions of the master integrals provides a dramatic simplification: in this limit the loop and phase space integrals factorize, making them much easier to evaluate. For example, expanding around $y=0$, Eq. (\ref{eq:2cut}) has the boundary condition
        \begin{multline}\label{eq:boundary}
	        \cI_1 = -\frac{ 2^{2 d-6} \pi ^2  \csc \left(\frac{\pi  d}{2}\right)}{(d-3)^2 \Gamma (d-3)}y^{2 d-6}\\
	         -\frac{ 2^{2d-7} \pi ^2  \csc \left(\frac{\pi  d}{2}\right) \Gamma \left(\frac{3-d}{2}\right)}{\Gamma \left(\frac{d-1}{2}\right)}y^{4 d-12}+\cdots	         
        \end{multline}
        The differential equations couple many different master integrals. Once they are solved using the boundary conditions, the individual masters may be extracted. For example, we find 
        \begin{multline}
            e^{2\epsilon \gamma_E}\cI_1
            =\frac{\pi  x}{(x+1) \epsilon }
            +\pi\bigg(
            \frac{(x-1)\ln(1+x)-3x\ln x+5x}{x+1}
            \\
            +\frac{\ln^2(1+x)+\Li_2(1-x^2)-\pi^2/6}{2x}
            \bigg)+\cO(\epsilon)
        \end{multline}
        We check the results for the master integrals by constructing their sums which determine the imaginary parts of the corresponding uncut diagrams and evaluating the latter numerically using \texttt{Fiesta} \cite{Smirnov2016}.
        More details of related calculations using the same technology can be found in~\cite{Lee:2020obg,Lee:2020zpo}.
        
    \end{fmffile}
    We use dimensional regularization $d=4-2\epsilon$ to regulate the infrared and ultraviolet divergences. The renormalization in the on-shell scheme amounts to the following relation between the renormalized and bare virtual corrections $\sigma_\text{virt}=\left(\sigma_\text{virt}\right)_\text{bare}+2\delta Z_{\psi} \sigma_\text{LO}+\delta\sigma_m$, where $\delta Z_{\psi}=-\frac{(4 \pi  \alpha ) (3-2 \epsilon ) \Gamma (\epsilon )}{(4 \pi )^{2-\epsilon } (1-2 \epsilon )}\left(\frac{e^{\gamma_E}}{4 \pi }\right)^{\epsilon }$ is the one-loop contribution to the wave function renormalization constant, $\sigma_\text{LO}$ is the Born cross section defined below in Eq. \eqref{eq:sigmaLO}, and $\delta\sigma_m$ is the sum of the last eight diagrams in Fig. \ref{fig:diagrams}, where cross denotes the one-loop mass counterterm $i\delta m=i m \frac{(4 \pi  \alpha ) (3-2 \epsilon ) \Gamma (\epsilon )}{(4 \pi )^{2-\epsilon } (1-2 \epsilon )}\left(\frac{e^{\gamma_E}}{4 \pi }\right)^{\epsilon }$ vertex (cf. Ref. \cite{Lee:2020zpo}). After the renormalization we can put $\epsilon=0$ in the sum $\sigma_\text{virt}+\sigma_\text{real}$ as the infrared divergences cancel between virtual and real contributions. 
    
    Let us write the cross section for $e^-\gamma \to e^-\gamma(+\gamma)$ to order $\alpha^3$ as $\sigma_\text{tot} = \sigma_\text{LO} + \sigma_{\text{NLO}}$ where
    \begin{multline}
    \label{eq:sigmaLO}
        \sigma_{\text{LO}} =\frac{\pi\,\alpha^2}{m^2(x+1)} \Big[
        \frac{x^3+18 x^2+32 x+16}{x^2 (x+1)}\\+\frac{\left(2 x^3-6 x^2-24 x-16\right) }{x^3}\ln (x+1) \Big]
    \end{multline}
    and
    \begin{widetext}
        \begin{multline}\label{eq:NLO}
            \sigma_{\text{NLO}} = \frac{\alpha^3}{m^2 x^3}\Big\{
            - \frac{x \left(273 x^3-982 x^2-2960 x-1744\right)}{24(x+1)^2}
            + \frac{ 37 x^4- 54 x^3 - 339 x^2- 428 x -184}{4(x+1)^2} \ln (x+1) \\
            + \frac{x^2 \left(14 x^4+17 x^3-17 x^2-22 x-8\right)}{2 (1 - x) (1 + x)^3} \ln x
            - \frac{4 x^6+35 x^5-31 x^4-755 x^3-1765 x^2-1506 x-440}{2 (x+1)^2 (x+4)}\ln^2 (x+1)
            \\
            - \frac{x^6+7 x^5-28 x^4-239 x^3-449 x^2-338 x-88}{(x+1)^2 (x+4)} \Li_2(-x)
            - \frac13\left(x^2-16 x-23\right)\ln^3 (x+1)
            + \left(x^2-x+2\right) \left[\Li_2(1-x)-\frac{\pi^2}{6}\right]
            \\
            + \frac{x^4+7 x^3+x^2-3 x-2}{(x+1)^2} \ln (x+1)\ln x
                        + \left(x^2+2 x-6\right)
            \Big[ \Li_3 (x^2) -  \Li_2 (x^2)  \ln x \Big]
            \\
            - \frac{4 \left(x^5+26 x^4+146 x^3+316 x^2+288 x+96\right)}{(x+1)^2 (x+4)} G(-2,-1;x)
            + 8 \left(x^2-4 x-6\right) G(-1,-2,-1;x)
            \\
            + 4 (2x^2 -x-6)G(-1,-1,0;x)
            + 2 \left(2 x^2-7 x-12\right)G(-1,0,-1;x)
            - (5 x^2+32 x-8) G(0,-1,-1;x)
            \\
            - 3 (x-2)(x+4)y\Big [G(0,y,-1;x) + 2 G(y,-1,0;x)\Big]
            - \frac{8 y \left(x^4+3 x^3-18 x^2-68 x-24\right)}{(x+4) x}G(y,0,-1;x)
            \\
            + \frac{3 y \left(5 x^4+14 x^3-96 x^2-352 x-128\right)}{(x+4) x}G(y,-1,-1;x)
            - \frac{16 y \left(x^4+2 x^3-24 x^2-80 x-48\right)}{(x+4) x}G(y,-2,-1;x)
            \\
            + \frac{3 x^4+18 x^3+44 x^2-8 x-64}{x} y G(y,-1;x)   
            - \frac{6 y \left(x^3-12 x-8\right)}{x} G(-4,y,-1;x)
            + 3 \left(3 x^2-8\right) G(y,y,-1;x)
            \Big\}
        \end{multline}
    \end{widetext}
    The $G$ functions are defined iteratively via
    \begin{equation}
        G(a,a_1,\ldots,a_n;x) = \int_0^x dw_a(x') G(a_1,\ldots,a_n;x'),
    \end{equation}
    where weights are 
    \begin{equation}
        dw_y(x)=\frac{y dx}{x},\quad dw_a(x)=\frac{dx}{x-a} \quad (a=-4,-2,-1,0)
    \end{equation}
    and $G(0;x)=\ln x$. When $y$ is not among the letters, these functions are conventional Goncharov polylogarithms as the notation suggests. The weight $dw_y$ can be rationalized by changing variables from $x$ to $y$ in Eq.~\eqref{xy}. Then one can express the $G$ functions as linear combinations of  $G(b_1,\ldots, b_n;y)$ with $b_k\in \left\{0,\pm 1,\pm i, \pm{i}/{\sqrt{3}}\right\}$, or, alternatively, as linear combinations of  $G(\ldots;z)$, with $z=\frac{1-y}{1+y}$ and indices being the 4th or 6th roots of unity.
   
    % All the functions appearing in $\sigma_\text{NLO}$ have explicit forms in terms of ordinary polylogarithms (see the appendix). 
    % % For example,
    % \begin{multline}
    %     G(-1,-2,-1;x) = 2 \Li_3(-1-x) \\-\left[\Li_2(-1-x)- \frac{\pi^2}{12}\right]\ln (x+1)+ \frac{3\zeta_3}{2}. 
    % \end{multline}
    Some of the integrals can be evaluated into relatively simple forms with an additional judicious variable change. For example, using $z=\frac{1-y}{1+y}=\frac{2+x-\sqrt{x(x+4)}}{2}$, we can compute
    \begin{multline}
        G(y,-1;x)=\int_0^x \frac{d x'}{\sqrt{x(x+4)}}\int_0^{x'} \frac{d x''}{x''+1}\\
        = -\int_1^z d \ln z'  \int_1^{z'} d\ln \frac{z''^3 + 1}{z''(z''+1)}\\
        =\frac{1}{3} \Li_2(-z^3) - \Li_2(-z)+ \frac{1}{2} \ln^2 z  - \frac{\pi^2}{18}
    \end{multline}
    In the Appendix we present the expressions for all the $G$ functions entering Eq. \eqref{eq:NLO} in terms of the classical polylogarithms $\Li_n$. The same expressions, as well as the total cross section, are given in a Mathematica notebook included with the arXiv submission of this paper. As a check, we compare to the numerical values given in Table 3 of~\cite{Denner:1998nk}, and find perfect agreement. 
    
\begin{figure}[t]
        \centering
        \includegraphics[width=0.9\columnwidth]{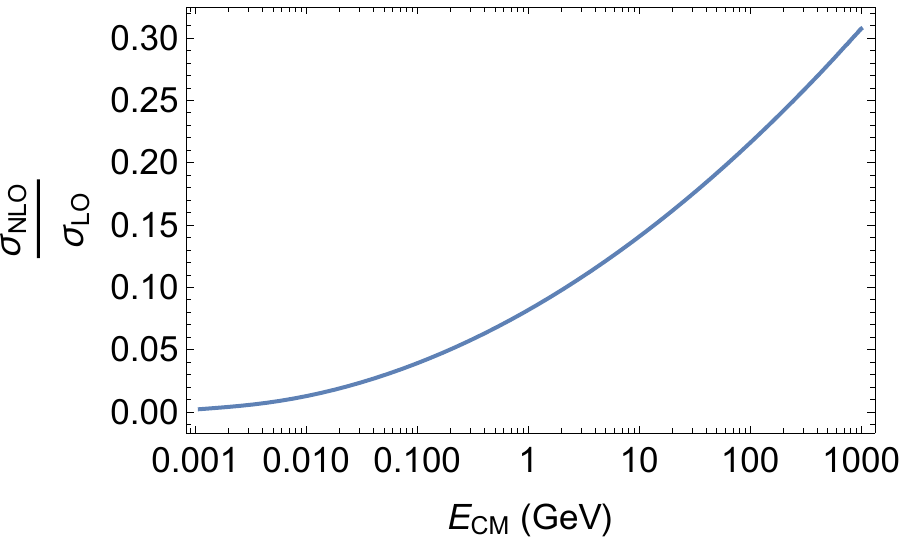}
        \caption{NLO corrections to the Compton scattering cross section, as compared to the leading order. At $E_\text{CM} = 1 \text{GeV}$ the correction is 8.2\% growing to a 30\% correction at 1 TeV. }
        \label{fig:plot}
    \end{figure}

    Figure~\ref{fig:plot} plots the size of the NLO correction relative to the leading order as a function of center-of-mass energy $E_\text{CM}= \sqrt{s}$, using $\alpha =1/137.036$.  At threshold $s\to m^2$, the NLO correction vanishes. This is guaranteed by Thirring's theorem~\cite{Thirring:1950cy,Dittmaier:1997dx,Denner:1998nk}, and is a non-trivial check on our computation. More precisely, using \texttt{PolyLogTools} \cite{Duhr:2019tlz}, we find that near threshold $(s\sim m^2)$, the cross section behaves as
    \begin{multline}
        \sigma_{\text{tot}} = \frac{\pi\alpha^2}{m^2} 
        \left[\frac{8}{3} -\frac{8}{3} x +\cdots\right] \\
        + \frac{\alpha^3}{m^2} x^2 \left[-\frac{16}{9}\ln x+\frac{7}{15}+\cdots\right]
    \end{multline}
    with the $\alpha^3$ term vanishing like $(s-m^2)^2$. 
    
    We see from Figure~\ref{fig:plot} that the NLO correction to the cross section grows with energy, providing a 30\% correction already at 1 TeV (pair production and electroweak corrections are also increasingly important at high energy~\cite{Dittmaier:1993bj}).  At high energy, the asymptotic behavior is
    \begin{align}
        &\sigma_{\text{tot}} = \frac{\pi\alpha^2}{s} 
        \left[2\ln \frac{s}{m^2}+1+\cdots \right] 
\\
&\nonumber
\!+\! \frac{\alpha^3}{s} \!\left[\frac{1}{3}\ln^3\frac{s}{m^2} \!-\!\frac{1}{2} \ln^2 \frac{s}{m^2}
\!+\!\frac{17}{4} \ln \frac{s}{m^2} \!-\!\frac{75}{8} 
\!-\!\frac{\pi^2}{2} +4\zeta_3\right]
    \end{align}
    This is a new result. The $\frac{1}{s} \ln^3\frac{s}{m^2}$ behavior of the NLO result dominates over the $\frac{1}{s} \ln \frac{s}{m^2}$ behavior at LO; this explains the growth of their ratio at high energy in Figure~\ref{fig:plot}.

    To summarize, we have calculated the Compton scattering total cross section with accuracy $O(\alpha^3)$. The result can be represented in terms of the Goncharov's polylogarithms of the  argument $z=\frac{\sqrt{s+3m^2}-\sqrt{s-m^2}}{\sqrt{s+3m^2}+\sqrt{s-m^2}}$ with letters involving $6$th and $4$th root of unity \footnote{If we want to avoid also the classical polylogarithms $\Li_n(x^2)$ and $\Li_n(1-x)$, we should include also the letters $\varphi^2$ and $\varphi^{-2}$, where $\varphi=\frac{1+\sqrt{5}}2$ is the golden ratio.},
    or in temrs of classical polylogarithms.
    % Alternatively, we can express the cross section in terms of classical polylogarithms $\Li_2$ and $\Li_3$ at the expense of lengthier expression. 
%    We also provided the leading terms in the expansion around the threshold and the  high-energy limits. 
The analytic form of the total cross section 
    %fills a hole in the existing literature. It
    may be of use in numerous applications, from cosmic ray astrophysics to particle colliders as well as to theoretical investigations into factorization and forward/backward scattering.
    
    Our analytic results reveal some remarkable features. For one, when continued to a multi-valued analytic function, the cross section manifests a branch point at $s=2m^2$, due to terms like $\Li_3(x^2)$. Although the singularity is on an unphysical sheet
    and the cross section is smooth around $s=2m^2$,  the presence of this branch point affects the threshold and high-energy expansions: both diverge from the exact result near $s=2m^2$ no matter how many terms one retains. Moreover, these divergences cannot be eliminated by any rational variable change. 
    %The same is true for the high-energy asymptotics below $s=2m^2$.
    This is in sharp contrast to the cross section of $e^-\gamma\to e^-e^+e^-$~\cite{Lee:2020obg} which is  nicely approximated after only a few terms of high-energy asymptotics, or a few terms of threshold asymptotics when expanding in $r=\frac{s-9m^2}{s-m^2}$.  One might hope to give a physical interpretation to the branch point by connecting sequential monodromies to multiple cuts, as was done at the amplitude level in~\cite{Bourjaily:2020wvq}.
    
    Another intriguing feature is the appearance of a double log $\ln\frac{s^2}{m^2}$ in the NLO cross section over the LO cross section. Normally, double logarithms are due to overlapping soft and collinear Sudakov singularities. However, the soft singularities are expected to cancel in QED cross sections due to the Block-Nordsieck theorem~\cite{Bloch:1937pw} leaving at most single logarithms at each order in $\alpha$. It would be interesting to understand the origin of this double logarithm, and its connection to DGLAP or BFKL evolution both of which are single logarithmic. 

    The authors would like to thank C. Vergu, H-X Zhu and Y. Zhu for discussions.
    %collaboration on early stages of this project.
    %and C. Vergu for discussions. 
    MDS is supported in part by the U.S. Department of Energy under grant DE-SC0013607. RNL is supported by  Russian Science Foundation under grant 20-12-00205. 
    
%    \newpage
 
 \appendix    
\begin{widetext}
    \section{Appendix: Explicit expressions in terms of classical polylogarithms} 
    \begin{align}
   	G&(-2,-1;x)=\Li_2(-x-1)
       +\ln (x+1) \ln(x+2)
       +\frac{\pi ^2}{12},\\
   	G&(y,-1;x)=\frac{\Li_2\left(-z^3\right)}{3}
       -\Li_2(-z)+\frac{1}{2}\ln ^2{z}
       -\frac{\pi ^2}{18},\\
   	G&(-1,-2,-1;x)=2 \Li_3(-x-1)
       -\Li_2(-x-1) \ln (x+1)
       +\frac{\pi ^2}{12}  \ln (x+1)
       +\frac{3 \zeta_3}{2},\\
   	G&(-1,-1,0;x)=\Li_3(-x)
       +\Li_3\left(\frac{x}{x+1}\right) 
       +\frac12\ln^2(x+1)\ln{x}
       -\frac{1}{6}\ln^3(x+1),\\
   	G&(-1,0,-1;x)=-2 \Li_3(-x)
       -2 \Li_3\left(\frac{x}{x+1}\right)
       +\Li_2(-x) \ln (x+1)
       +\frac13\ln ^3(x+1),\\
   	G&(0,-1,-1;x)=\Li_3(-x)
       +\Li_3\left(\frac{x}{x+1}\right)
       -\Li_2(-x) \ln (x+1)
       -\frac{1}{6} \ln ^3(x+1),\\
   	G&(0,y,-1;x)=
    	2 \Re\left[
    	   2\Li_3\left(\frac{1-z}{1-r\,z}\right)
    	   -2\Li_3\left(\frac{z-1}{z-r}\right)
    	   +\Li_3\left(1-r/z\right)
    	   -\Li_3\left(1-z/r\right)
        \right]
        +2 \Li_3\left(1-z^{-1}\right)
    	-2 \Li_3(1-z)\nonumber\\&
    	+\left[
            \Li_2(1-z)
            -\Li_2\left(1-z^{-1}\right)
        \right] \ln \left(x+1\right)
    	+\arctan\left(\sqrt{3}\tfrac{1-z}{1+z}\right) \left[
            \frac{2\pi}{3} \ln \left(x+1\right)
            +\frac{2\left(\psi ^{\prime}\left(\tfrac{1}{6}\right)-2\pi^2\right)}{5 \sqrt{3}}
        \right]
    	-\frac{1}{3} \ln ^3{z}
    	+\frac{\pi ^2}{18} \ln {z},\\
   	G&(y,0,-1;x)=
    	2 \Re\left[
    	   2 \Li_3\left(r-r^2/z\right)
    	   -2 \Li_3\left(r-r^2z\right)
    	   +\Li_3\left(1-z/r\right)
    	   -\Li_3\left(1-r/z\right)
    	\right]
    	+\big[
           \Li_2\left(1-z^{-1}\right)
           -\Li_2(1-z)\nonumber\\&
    	   -\ln{x} \ln{z}
       \big]\ln \left(x+1\right)
    	-\arctan\left(\sqrt{3}\tfrac{1-z}{1+z}\right) \left[
    	   2 \pi  \ln{z}
           +\frac{2}{5 \sqrt{3}}\left(\psi ^{\prime}\left(\tfrac{1}{6}\right)-2\pi^2\right)
       \right]
    	+\frac{11\pi ^2}{18}  \ln {z},\\
   	G&(y,-1,0;x)=
    	\frac49 \Li_3\left(-z^3\right)
    	+\frac13 \Li_2\left(-z^3\right) \ln {x}
    	-2 \Li_3\left(z-z^2\right)
    	-2 \Li_3(z-1)
    	+\frac29 \Li_3\left((z-1)^3\right)
    	-4 \Li_3(-z)
    	\nonumber\\&
    	-2\Li_3(z)
    	-\Li_2(-z) \ln {x}
    	-\frac16 \ln ^3{z}
    	+\frac16 \pi ^2 \ln{z}
    	-\frac19 \pi ^2 \ln (1-z)
    	-\frac23 \zeta_3,\\
   	G&(-4,y,-1;x)=
    	2 \Re\bigg[
            \Li_3\left(1+z^{-1}\right)
            -\Li_3(1+z)
            +\Li_3\left(1-r/z\right)
            -\Li_3\left(1-z/r\right)
            +2 \Li_3\left(\frac{1-z/r}{1+z}\right)
            -2 \Li_3\left(\frac{z-r}{z+1}\right)
        \bigg]\nonumber\\&
    	-\left[
            \Li_2(-z)
            -\Li_2\left(-z^{-1}\right)
            +\ln \left(x+4\right) \ln{z}
        \right]\ln \left(x+1\right)
    	-\frac{2}{3} \arctan\left(\sqrt{3}\tfrac{1-z}{1+z}\right) \left[
            \frac{2 }{5 \sqrt{3}}\left(\psi ^{\prime}\left(\tfrac{1}{6}\right)-2\pi^2\right)
            -\pi  \ln \left(x+4\right)
        \right]\nonumber\\&
    	-\frac{1}{3} \ln ^3{z}+\frac{5}{9} \pi ^2 \ln{z},\\
   	G&(y,-2,-1;x)=2 \Re\bigg[
            \Li_3\left(\frac{1 - r/z}{1 - i\,r}\right)
            +\Li_3\left(\frac{1 - r/z}{1 + i\, r}\right)
            -\Li_3\left(\frac{1-z/r}{1+i/r}\right)
            -\Li_3\left(\frac{1-z/r}{1-i/r}\right)
            -\Li_3\left(1-r/z\right)\nonumber\\&
            +\Li_3\left(1-z/r\right)
            +\frac{1}{2} \ln \left[(1-z/r)(1-r/z)\right]
            \Big[
                \Li_2\left(\frac{1-z/r}{1+i/r}\right)
                +\Li_2\left(\frac{1-z/r}{1-i/r}\right)
                -\Li_2\left(\frac{1 - r/z}{1 - i\,r}\right)\nonumber\\&
                -\Li_2\left(\frac{1 - r/z}{1 + i\, r}\right)
                +\Li_2\left(1-r/z\right)
                -\Li_2\left(1-z/r\right)
            \Big] 
    	\bigg]
        -\frac{1}{36} \left(5 \pi ^2+18 \ln ^2(2-\sqrt{3})\right) \ln{z},\\
    G&(y,-1,-1;x)=
        2 \Re\bigg[
            \Li_3\left(1-z/r\right)
            -\Li_3\left(1-{r}/{z}\right)
            +\Li_3\left(\frac{1-z/r}{\sqrt{3}i z}\right)
            -\Li_3\left(\frac{r-z}{\sqrt{3}i}\right)\nonumber\\&
            -\frac12\ln \left(x+1\right) \left[
                \Li_2\left({r}/{z}\right)
                -\Li_2\left(r\, z\right)
            \right]
        \bigg]
        +\frac{1}{12} \ln {z} \left(
            \ln ^2{z}
            +3 \ln ^2\left[3\left(x+1\right)\right]
            -6 \ln ^2{3}
            +\pi ^2
        \right)\nonumber\\&
        -\left[\arctan\left(\sqrt{3}\tfrac{1-z}{1+z}\right)\right]^2\ln{z}
        -\frac{1}{15}\arctan\left(\sqrt{3}\tfrac{1-z}{1+z}\right) \bigg[
            5 \pi  \ln \left[3\left(x+1\right)\right]
            +\frac{2 }{\sqrt{3}}\left(\psi ^{\prime}\left(\tfrac{1}{6}\right)-2\pi^2\right)
        \bigg],\\
    G&(y,y,-1;x)=\Li_3(-z)-\frac19\Li_3\left(-z^3\right)-\frac16 \ln ^3{z}+\frac{\pi ^2}{18} \ln {z}+\frac23 \zeta_3.
    \end{align}
Here $\psi^{\prime}(x)$ is the derivative of the digamma function $\psi(x)=\Gamma^{\prime}(x)/\Gamma(x)$ and $r=e^{i\pi/3}$ is the primitive sixth root of unity.
\end{widetext}       
    \bibliography{refs}

\end{document}